\documentclass[copyright,creativecommons]{eptcs}
\providecommand{\event}{FOCLASA 2014} 
\usepackage{breakurl}             
\usepackage{graphicx}
\usepackage{stmaryrd}
 \usepackage{mathtools,mathpartir}
\usepackage{amsmath,amssymb,amscd,mathrsfs}
\usepackage{epsfig,color,subfigure,enumitem}

\setlist{noitemsep,topsep=.5ex}
\usepackage{macrospNets}
\usepackage[english]{babel}
\newcommand{\Nature}{\symb{Nature}}

\newcommand{\Membrane}{\symb{Membr}}
\newcommand{\Content}{\symb{Cont}}
\newcommand{\Symmetry}{\symb{Sym}}
\newcommand{\ControlLevel}{\symb{CL}}

\title{Verifying the correct composition of distributed components: Formalisation and Tool
\thanks{This work was partialy funded by the Associated Team DAESD
  between INRIA and ECNU, Shanghai; and by the french {\em Fond National pour la Soci\'et\'e Num\'erique} project OpenCloudware.}}
\author{Ludovic Henrio 
\institute{Univ. of Nice Sophia Antipolis, CNRS, France}
\email{ludovic.henrio@cnrs.fr}
\and
Oleksandra Kulankhina
\institute{INRIA Sophia Antipolis M\'edit\'erann\'ee, France}
\institute{Univ. of Nice Sophia Antipolis, CNRS}
\email{oleksandra.kulankhina@inria.fr}
\and
Dongqian Liu 
\institute{MoE Engineering Research Center for Software/Hardware}
\institute{Co-design Technology and Application, ECNU Shanghai, China}
\email{maggie.liudon@gmail.com}
\and
Eric Madelaine 
\institute{INRIA Sophia Antipolis M\'edit\'erann\'ee, France}
\institute{Univ. of Nice Sophia Antipolis, CNRS}
\email{eric.madelaine@inria.fr}
}
\def\titlerunning{Verifying the correct composition of distributed components: Formalisation and Tool}
\def\authorrunning{L. Henrio, O. Kulankhina, D. Liu, E. Madelaine}

\begin{document}

\maketitle


\begin{abstract}
This article provides formal definitions characterizing
well-formed composition of components in order to guarantee their safe
deployment and execution. Our work focuses  on the
structural aspects of component composition; it puts together most of the
concepts common to many component models, but
never formalized as a whole. Our formalization characterizes correct
component architectures made of functional and non-functional aspects,
both structured as component assemblies. Interceptor chains can be used for a safe and
controlled interaction between the two aspects.
Our well-formed components guarantee a set of properties ensuring that
the deployed component system has a correct architecture and can run
safely. 
Finally, those definitions constitute the formal basis for our Eclipse-based environment for the
development and specification of component-based applications.
\end{abstract}

\section{Introduction}
Designing safe distributed component-based software systems is a
challenging task including a wide range of issues ranging from the
safe composition of components to the applications performance and
reliability. This work focuses on the correctness of component
 assembly which is a strong prerequisite for correct deployment and
execution of components.  We target component models which are
hierarchical~\cite{Sofa2Reconfig-HP06} and provide runtime control on the
component application.  In component-based software engineering, this
last feature, i.e. the control and management of the system
architecture and of the business application, is part of what is generally 
called \emph{non-functional} concern. 
More precisely, non-functional aspects may cover many aspects, from security,
latency, time aspects, energy or cost, to management concerns including 
mechanisms for dynamic
reconfiguration, fault-tolerance, load balancing, self-healing, or more
elaborate autonomic behaviours. In this paper, we are essentialy considering
the second category, i.e. the management aspects, and we want to support the ability to design
non-functional aspects themselves as a component system, which ensures a clear
design of the whole application. 
The business logic of the application is called the \emph{functional aspect}.
 
Our main goal  is to provide a clear and
formal definition of component assemblies and a set of logic
predicates ensuring the correctness of component composition and
component interconnection. 
In particular our predicates ensure a strong separation of
concerns: a non-functional component can only interact with
non-functional components, or with non-functional interfaces of
business components. This way, each binding between components only
transmits either functional or non-functional invocations.
However a clean interaction between these concerns is highly desirable:
we introduce
interceptors, which are special components providing interaction between
functional and non-functional elements. They can observe
functional invocations and trigger a reaction of the control part of
the component. In the other direction, non-functional components can
influence the functional behavior by modifying its behaviour in two
ways: either through reconfiguration, i.e. modification at runtime of
the component architecture, or by changing parameters, i.e. component
attributes, that will
influence the functional behaviour.


In this paper we focus on one specific component model: the
Grid Component Model (GCM)~\cite{BCDGHP:article2009}.  GCM allows the
construction of distributed hierarchical applications with
asynchronous group communications and strong separation between
business logic and control parts of a system. To our knowledge, GCM is 
one of the only established component models that provides specific features 
for managing complex software systems, and that offers  architectural artifacts
and methodological approaches to organise the control part of such applications.
 GCM-based systems can
be designed and modeled using the VerCors tool\footnote{The VERCORS
  platform: VERification of models for distributed communicating
  COmponants, with safety and Security.
  \url{http://team.inria.fr/scale/software/vercors/}};
An XML-based {\em Architecture Description Language} (ADL) file, 
describing the architecture of the application, can then be
generated by VerCors and used to build and deploy the GCM application, using
the GCM/ProActive execution environment~\cite{BHR:SPE14}.
We started from the formal representation of GCM architecture given in 
\cite{ameurboulifa:hal-00761073} and introduced a number of
important concepts. This article has the following new contributions:
\begin{itemize}
\item
First, we provide a formal representation of component architectures including a flexible set of constructs for the definition of non-functional part of an application which has never been formalized before. The business logic and control parts of a system are strongly separated.
\item
Second, we formalize the notion of interceptors and define a number of auxiliary predicates insuring their safe composition with the rest of the system. 
\item Third, we specify a set of key properties for component
  architecture, those properties are guaranteed for any component
  assembly that is well-formed according to the rules defined in
  Section~\ref{sec-wf}.  After validation of these rules, we garantee
  that the generated ADL code is correct, in the sense that building
  the component-based application by running the GCM component factory
  will terminate successfully with no runtime error. Additionally, the
  properties we exhibit are necessary prerequisite for the correct
  execution and the reconfigurability of the system; those properties
  include uniqueness of naming, separation of concerns, and
  communication determinacy.
\item
Finally, we implement an environment allowing graphical modeling and validity check for 
the distributed systems architecture with respect to the proposed formalization. 
\end{itemize}

The purpose of our formalization is threefold. First it is component
model independent: by formalizing the correctness rules, we make them
applicable outside the scope of GCM, and of our tools.  Second, it
makes it easier to re-use the formalization for different purposes,
e.g. to implement different checkers, or encode it in a theorem
prover.  Finally, it allows reasoning on the properties ensured by the
component model: properties of Section~\ref{sec:properties} can
be formally proven from the definition of component correctness.

This paper is organized as follows. Section~\ref{sec:pnets-def}
describes the Grid Component Model and our modeling
platform. Section~\ref{sec:formal} presents the  
formalization of GCM architecture and of correct architecture. 
Section \ref{sec:rw} describes how to adapt our formalism to other
features or other component models and
presents an overview of the related work.
 Section \ref{sec:vce} describes our
environment for modeling GCM-based systems.

\section{Context}\label{sec:pnets-def}

\subsection{GCM}

GCM~\cite{BCDGHP:article2009} is a component model targeting
large-scale distributed applications. It is an extension of the
Fractal~\cite{fractal:SPE2006} component model; we summarize below the
architectural aspects of GCM, and its main advantages. We also provide an illustrative example of a real-world application built using GCM.


\begin{figure}[t]
     \centering
     \includegraphics[width=15cm]{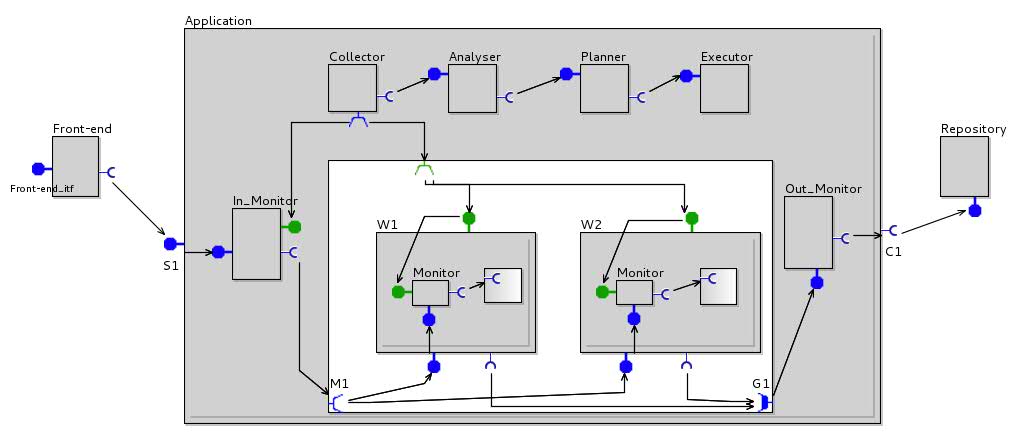}
     \caption{A typical GCM component assembly (image produced by VerCors)}
     \label{fig:gcm}
 \end{figure}


A GCM component application is a composition of components, interfaces and
bindings. 
GCM is a hierarchical component model, consequently
there are two types of components in GCM: primitive and composite
ones. Composite components are used to compose other components while
the primitive ones
are the leaves of the composition hierarchy.
More precisely, a composite component (\textbf{Application} in Figure~\ref{fig:gcm})
 is made of two parts: membrane (gray
part) and content (white part). The content contains the 
sub-components dealing with the functional aspects. The membrane
includes sub-components which are responsible for everything besides
business logic: control, monitoring, structural
reconfiguration, etc. 
 A primitive component is similar except that the content is not known (e.g., \textbf{Front-end}, \textbf{Repository} and \textbf{W1} in Figure~\ref{fig:gcm}). It
encapsulates some business code. It has a list of methods, which it
can execute.

The communication between components is performed in the form of
method invocation. In Fractal and GCM interfaces are similar to object
interfaces characterized by a set of methods that can transit through
this interface. There are client and server
interfaces. Client interfaces (e.g. \textbf{C1}) emit the methods invocations. Server
interfaces (e.g. \textbf{S1}) receive them. We also distinguish the interfaces dealing
with the business logic (\emph{functional interfaces} drawn in blue color) and the ones dealing with the control of an
application (\emph{non-functional} ones colored in green). 
A binding is an arrow from a client to a
server interface, communications between components follow those bindings.
\begin{figure}[t]
     \centering
     \includegraphics[width=8cm]{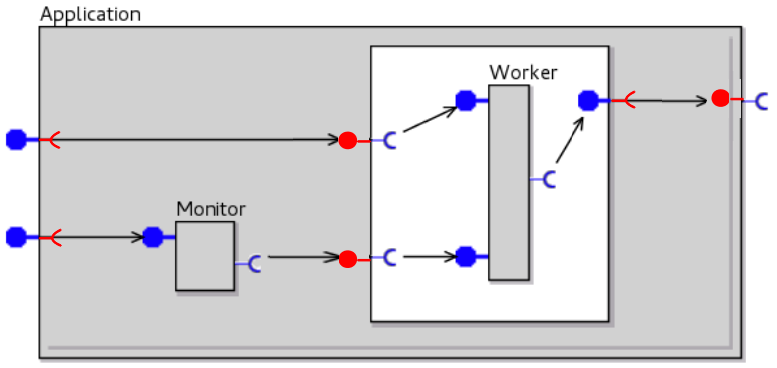}
     \caption{Internal interfaces of a membrane}
     \label{fig:membr_itf}
 \end{figure}

 In GCM, the interfaces can be attached to the components.  An
 interface that is accessible from the exterior of a component is
 called an \emph{external interfaces} (e.g. \textbf{S1}).  When an
 interface is reachable from the inside of a component, i.e. from its
 content, it is called an \emph{internal interfaces} (e.g.
 \textbf{M1}). The membrane of a component can also have internal
 interfaces, but they are not declared explicitly. The set of internal
 interfaces of a membrane is represented by the union of two other
 sets: all the external interfaces of the component containing the
 membrane and all the internal interfaces of the content inside the
 component. All the interfaces are taken with the same properties, but
 with an opposite role (e.g. a server interface becomes a client one
 and vice-verse). Figure \ref{fig:membr_itf} illustrates the (implicite) 
internal interfaces of a membrane, displayed in red color.

Figure~\ref{fig:gcm} shows an illustrative example taken
from~\cite{BHR:SPE14}. It represents the architecture of a real-world
GCM system implementing an autonomic master-slave application
treating incoming requests in parallel. Component \textbf{Front-end} is a
front-end of the application. It receives the job-requests from the
user. The requests are, then, processed by the workers (\textbf{W1}
and \textbf{W2}). The results are sent to the repository represented
by a component \textbf{Repository}.
The application reconfigures itself autonomically: the workers are
added or removed depending on the system workload. This is realized by
the sequence of components: \textbf{Collector}, \textbf{Analyser},
\textbf{Planner}, and \textbf{Executor}. This sequence follows the
principle of the MAPE model for autonomic computing. The Monitoring
components (\textbf{Monitor}, \textbf{In\_Monitor}) in the membranes of the workers and in \textbf{Application} monitor
different metrics such as served requests per
minute, and available disk space, CPU, and memory. Component \textbf{Collector} gathers the information from
the Monitoring components. It sends the information to the 
component \textbf{Analyser}. Based on the received information,
\textbf{Analyser} sends an alarm to the Planning component (\textbf{Planner}).
The former one plans the reconfiguration, which is enacted by the Execution
component (\textbf{Executor}).

GCM provides the following features to the applications:
\begin{description}
\item[Strong separation of concerns] The functional part of a GCM
  application can be separated from the control part thanks to the
  separation of a composite component into a membrane and a content
  and to the usage of functional and non-functional interfaces. In the
  provided example, the components performing the reconfiguration
  (\textbf{Collector, Analyser, Planner, Executor}) are separated from the ones responsible
  for the business logic (\textbf{W1, W2, Repository}). This separation
  ensures, as much as possible, the independence of the business code -- the request treatment
  in the example -- from the management code -- the sequence of control components in the example.

\item[Reconfiguration] In GCM, the architecture of the application can
  be accessed and modified at runtime; this is realized in the example by the execution component. It allows the
  non-functional concern to modify the structure of the functional (and the
  non-functional) part of the component application, and thus to
  change its behavior.
 
\item[Interceptors] Sometimes, some information should be shared
  between the functional part and the controllers of the application.
  Interceptors are specific components inside the membrane that can
  intercept the flow of functional calls in order to trigger reaction
  from the non-functional aspects.  For example, in the use-case,
  \textbf{Monitor} components are monitoring the number of
  job-requests sent to the worker, they illustrate very well what is
  an interceptor. In the membrane, functional
  interfaces are directly connected, 
  however one or several interceptors can be inserted within such a
  binding and interact with other non-functional components (see the
  non-functional binding between \textbf{Monitor} and \textbf{Collector} in
  the example).
  
\item[One-to-many / many-to-one communications]  GCM also targets 
large-scale distributed systems where an invocation is to be
broadcast to several entities or simply sent to different servers to
ensure load-balancing. For this, GCM introduces one-to-many client
interfaces, called multicast interfaces like the interface \textbf{M1} of
Figure~\ref{fig:gcm}. From a structural point of view the
particularity of multicast interfaces is that they can be connected
to several server interfaces. Symmetrically, GCM introduce gathercast server
interfaces for synchronizing many-to-one communications, typically
waiting for a number of communications before triggering a
communication. 
\end{description}


Our main contribution here is a formal definition
and specification of the non-functional aspects of components; we will also
see how we formally insure the separation of concerns and how we formally specify the interceptors.

\begin{figure}[ht]
     \centering
     \includegraphics[width=.8\textwidth]{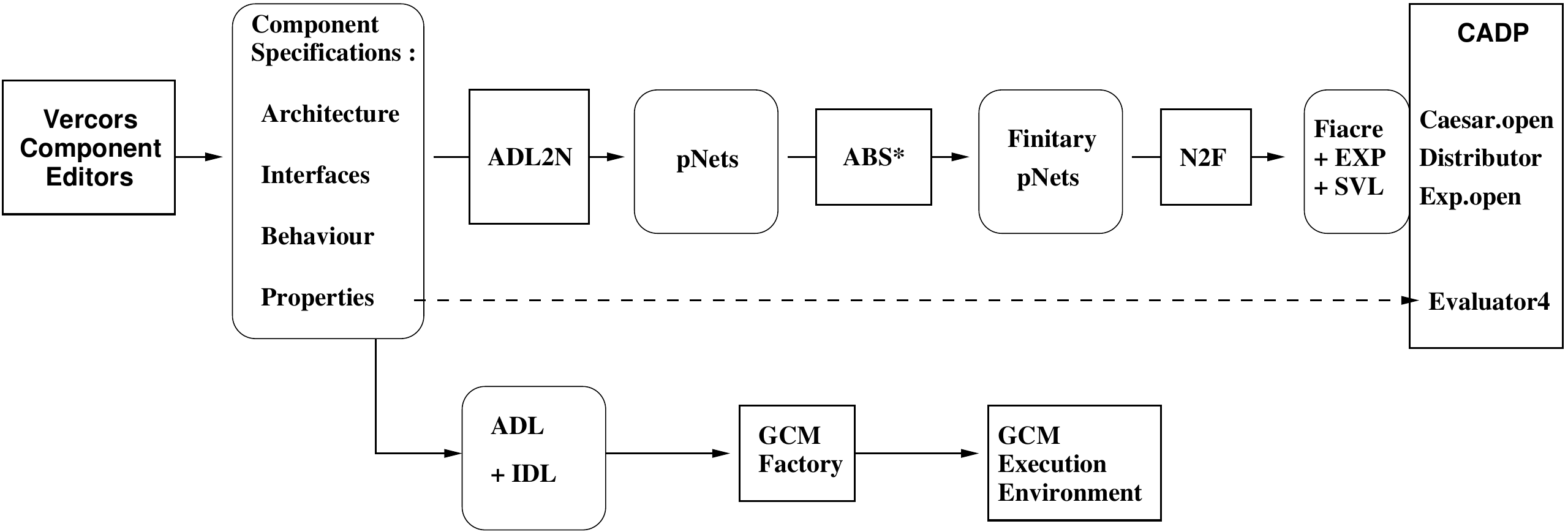}
     \caption{Principles of the VerCors platform}
     \label{fig:vercors_structure}
 \end{figure}

\subsection{VerCors}
VerCors is a platform for the specification, analysis, verification,
and validation of the GCM-based applications. The principle of the
tool is illustrated in Figure \ref{fig:vercors_structure}. First, a
user specifies the architecture of a GCM-based application, the signature of its interfaces, and the
behavior of the primitive components using VerCors Component Editor
(VCE). Then, from these descriptions the behavioral model of an
application is generated (ADL2N) in a form of a dedicated behavioral semantic model:
a parameterized networks of synchronized automata (pNets). 
This semantic model is transformed by abstraction functions (ABS*), until reaching a
finite model suitable for model-checking. Finally, the
resulting behavioral model is translated (N2F) into a set of languages
used by a Model Checker (CADP tool \cite{CADP-Tacas2011}). Finally, the Model Checker
verifies the correctness of the model with respect to a set of  
temporal logic properties (user requirements) and in the case of detected errors it
provides their description. 

Once the requirements have been proven correct on the VCE specification, the user can generate the set of files (ADL for the Architecture Description, IDL for the Interface definition in the form of Java interfaces). Then naturally, user has to provide java classes implementing the service methods of the primitive components.
These files are processed by the GCM component factory to build an executable application, that is executed with the GCM/ProActive execution environment.

In this paper we focus only on the structural aspect of GCM-based
applications which is covered by the VerCors Component Editor. 
Our purpose is to ensure that the components provided to the rest of
the tool chain are well-formed; for this we describe below a set of
rules that must be satisfied to ensure the correctness of a GCM
component assembly, we  describe their implementation in
VCE in Section~\ref{sec:vce}.

Using VerCors tool, we were able to design the architecture of the example illustrated at Figure~\ref{fig:gcm}, check its static semantic correctness properties, generate ADL files and deploy the corresponding application. 
\section{Formalization}
\label{sec:formal}
The purpose of this section is to define formally well-formed
components and exhibit their properties.
\begin{table}
\begin{center}
  \begin{tabular}{ p{3 cm} | p {10 cm} }
 \hline
 
\multicolumn{1}{c|}{Element} & \multicolumn{1}{c}{Formalization} \\ 
 \hline
 
 Server interface &
 $\symb{SItf} ::= (\textsf{Name}, \symb{MSignature}_i^{i\in I}, Nature)_S 		 \label{eq:SItf_def}  $\\ 
 \hline
 
 Client interface &
 $\symb{CItf}::= (\textsf{Name},  \symb{MSignature}_i^{i\in I}, Nature)_C		 
 \label{eq:CItf_def} $\\
 \hline
 
 Interface &
 $\Itf ::= \symb{SItf} ~|~ \symb{CItf}		 
 \label{eq:Itf_def} $\\
 \hline

 Method signature &
 $\symb{MSignature} ::= \textsf{MName} ~\times~ \textsf{Type} ~\to~ \textsf{Type} $\\
 \hline
  
 Primitive &
 $\symb{Prim} ::= \textsf{CName} <\symb{SItf}_i^{i\in I}, \symb{CItf}_j^{j\in J}, M_k^{k\in K}, \Membrane> $\\
 \hline

 Composite &
 $\symb{Compos} ::= \textsf{CName} <\symb{SItf}_i^{i\in I}, \symb{CItf}_j^{j\in J}, \Membrane, \Content> $\\
 \hline
  
 Component &
 $\symb{Comp} ::= \symb{Prim} ~|~ \symb{Compos} $\\
 \hline
 
 Content &
 $\Content ::= <\symb{SItf}_i^{i\in I}, \symb{CItf}_j^{j\in J}, \symb{Comp}_k^{k\in K},     \symb{Binding}_l^{l\in L}>$ \\
 \hline
 
 Binding &
 $\symb{Binding} ::= (\symb{QName}, \symb{QName}) $ \newline
 $\symb{QName}  ::= \textsf{This}.\textsf{Name}~|~\textsf{CName}.\text{Name} $\\ 
 \hline
  
 \hline
 
 
 
 Nature &
 $\symb\Nature ::= \textsf{F} ~|~ \textsf{NF} $ \\
 \hline
 
 Membrane &
 $\Membrane ::= <\symb{Comp}_k^{k\in K},  \symb{Binding}_l^{l\in L}>$ \\
 \hline
  
  \end{tabular}
\end{center}
\caption{The formalization of GCM architecture}
\label{table-ext}
\label{table-core}
\end{table}
\subsection{Structure}\label{sec:structure}
We start with the formal model representing the GCM architecture. First, we discuss
the core elements. Second, we formalize the
non-functional part of the GCM architecture. 

\subsubsection{Core}
In this section we define the core elements of GCM,
namely: Interfaces, Components and Bindings. Their formalization is
given in Table \ref{table-core}. We denote $A_i^{i\in I}$ a set of elements $A_i$ indexed in a set $I$. The formal definition is not provided for some elements of the table, because they are the terminal symbols. Such elements are presented in different font (e.g. \textsf{Name}, \textsf{Type}, \textsf{NF}).

Each server (\symb{SItf}) or client (\symb{CItf}) interface is
characterized by three attributes: \emph{Name} is the name of an
interface; \emph{MSignatures}
represent the methods which can be served by a server interface or
called by a client interface, with their signature; \emph{Nature}
defines if it is a functional or a non-functional interface. 

Each component is characterized by its name, the sets of external
client and server interfaces, and its membrane. A primitive component
also includes a set of local methods. A composite component includes
its content with a specific structure. The membrane is responsible for
the non-functional activity. The content is defined by the sets of internal client and
server interface, sub-components and a set of bindings located inside
the content.  

A binding connects two interfaces. It is described as a couple of
names defining its source and target interfaces. Each name consists of
two parts. The first part describes the container of the interface,
either the name of a sub-component or the identifier \emph{This}. The
second part is the name of the interface itself.

\subsubsection{Non-functional aspects}
The $Nature$ property of interfaces can be either \emph{functional
  (F)} or \emph{non-functional (NF)}.  The membrane is the part of a
component which is responsible for the non-functional aspect of an
application. Its formalization is given in the bottom of Table
\ref{table-ext}. The bindings ($\symb{Binding}_l^{l\in L}$) are the
connections between the interfaces in the membrane. A membrane can
contain a set of sub-components ($\symb{Comp}_k^{k\in K}$). The
interceptors are recognized in this set essentially from their binding
patterns, all the other components in a membrane are simple
controllers.
\subsection{Auxiliary functions}
In this section we introduce the auxiliary functions used for the GCM architecture well-formness formalization. The  auxiliary functions are given in Table \ref{table-aux}.

\begin{table}
\begin{center}
  \begin{tabular}{ l | p {13 cm} }
 \hline
 
 \multicolumn{1}{c|}{Function name} & \multicolumn{1}{c}{Function Definition} \\ 
 \hline
 
  Sym &
  $\Symmetry : \Itf \to \Itf$ \\
  \hline
 
\Itf &
$GetItf: (\Membrane ~|~ \Content ~|~ \Comp) ~\to~ \symb{Itf}$ \newline
$GetItf(X : Comp ~|~ Cont) ::= SItf(X) \cup CItf(X) $ \newline
$GetItf(X : Membr) ::= \Symmetry \bigg( \Itf(Parent(X))\cup \Itf(Cont(Parent(X)) \bigg) $ 
\\ \hline
 
 GetSrc &
$GetSrc: Binding ~\times~ (\Membrane ~|~ \Content) ~\to~ \symb{Itf}$ \newline
$GetSrc((Src, Dst), ctnr)::=$ 
$\symb{itf} ~\text{s.t.}~ Name(\symb{itf}) = Name (Src) \land $ \newline
$\left\{\begin{array}{l} 
\symb{itf} \in \Itf(ctnr) \land Role(\symb{itf}) = \textsf{C}, ~if ~Src = \textsf{This.Name} \\ \nonumber
\symb{itf} \in \Itf(comp).(comp \in ctnr \land \symb{Name}(comp)=\textsf{CName}),if  ~Src = \textsf{CName.Name}
\end{array}\right.
$
\\
 \hline

 GetDst &
$GetDst: Binding ~\times~ (\Membrane ~|~ \Content) ~\to~ \symb{Itf}$ \newline
$GetDst((Src, Dst), ctnr)::=$ 
$\symb{itf} ~\text{s.t.}~ Name(\symb{itf}) = Name (Dst)\land  $ \newline
$\left\{\begin{array}{l} 
\symb{itf} \in \Itf(ctnr) \land Role(\symb{itf}) = \textsf{S}, ~if ~Dst = \textsf{This.Name} \\ \nonumber
\symb{itf} \in \Itf(comp).(comp \in ctnr \land Name(comp)=\textsf{CName}),if  ~Dst = \textsf{CName.Name}
\end{array}\right.
$
 \\
 \hline
 
 Parent &
 $Parent :  (\symb{Itf} ~|~ \symb{Comp} ~|~ \Membrane ~|~ \Content) ~\to~ \Membrane ~|~ \Content ~|~ \symb{Comp}$ \\
 \hline 
  
  
  \end{tabular}
\end{center}
\caption{Auxiliary functions}
\label{table-aux}
\end{table}

First, we use auxiliary functions providing the access to the attributes of the interfaces and components. For example, the $Nature(\symb{itf}:\Itf)$ function returns the nature of an interface, $Binding(membr:Membr)$ returns the set of bindings in a membrane. The definitions of such functions are straightforward and we omit them in Table \ref{table-aux}. 

Second, the symmetry function ($Sym$) takes an interface as an input
and returns an interface with exactly the same properties but with an
opposite role: a client $(\ldots)_C$ interface becomes a server $(\ldots)_S$ one
and symmetrically. to 

Third, we introduce a \textit{GetItf} function. It takes a container (a
component, a membrane or a content) as an input and computes the set
of interfaces stored by it. If its argument is a component or a
content, then its result is the union of its  server and client
interfaces sets. 
The case of membrane is more complicated as its \emph{internal
  interfaces} are not explicitly defined. Instead, we compute the
set of the symmetric of the interfaces belonging to the component
containing the membrane and its content.

Finally, most of the consistency rules dealing with the bindings will use the auxiliary functions $GetSrc$ and $GetDst$. They are able to retrieve the Interface objects which represent respectively the source and the destination ends of a binding. If $GetSrc$ and $GetDst$ functions are applied to a binding inside a membrane, then they may return an internal interface of the membrane. 

Conversely, some rules use the $Parent$ auxiliary function, that
recovers the container (component, membrane or content) of an interface, a component, a content or a membrane.



\subsection{Interceptors}\label{sec:interceptors}
In this section we define interceptors more precisely. First, we provide the
non-formal definition of an interceptor. Then, we introduce a predicate
recognizing the interceptors among the other sub-components of a
membrane. 

An interceptor is a functional component inside a membrane. It is
connected to one external and one internal functional interfaces of the
membrane's parent. 
An
interceptor "intercepts" a functional call that goes from outside to
inside of the membrane (input interceptor) or vice versa (output
interceptors). The interceptors are used to monitor an application and
its functional calls. The only functional activity of an interceptor
should be to 
intercept and forward the functional calls through its functional
client and server interfaces. All the other actions are
performed through non-functional interfaces. To allow more
modularity in the design, interceptors can be assembled in
\emph{chains} inside the membrane. Interceptors in a chain must all be
either input or all output; we shall speak of input chains
and output chains.

In principle, it would be possible to relax this definition, and allow for more
general interceptor structures, e.g. including some parallelism in the form of multicast
client interface in the ``chain'', allowing more efficient processing. However it is
not clear whether this would be useful fo rreal applications, and this not implemeted 
in the GCM/ProActive middleware, so we prefer to keep the (relatively) simpler form 
in the formal definition.

In order to distinguish formally the interceptors from the other components, we
define a predicate $\symb{IsInterc}$. It is given in Table\ref{table-Interc-pred}. It takes a component and a membrane as
an input and returns $true$ if the component belongs to a chain of
interceptors inside the given membrane. An interceptor chain consist
of one or several interceptors. $\symb{IsInterc}$ uses a predicate $\symb{IsIntercChain}$ which identifies if a given sequence of $K$ components is a chain of pipelined interceptors inside the given membrane. $IsIntercChain$ predicate checks the following features of the indexed set of components given as input:
\begin{itemize}
\item
all the components are nested inside the membrane;
\item
all the components have exactly one functional server and one functional client interface;
\item
a functional call can go through the sequence of components. More formally, for any $k \in {2...K}$ there is a binding connecting client functional interface of the component number $k-1$ and server functional interface of the component number $k$;
\item   
the first component of an input chain intercepts a functional call to
the content while the last component forwards the
functional call to the content or vice versa for an output chain.
\end{itemize}

Predicate $\symb{IsIntercChain}$ uses two auxiliary functions: $\symb{GetSItf}_F(\symb{comp})$
(resp. $\symb{GetCItf}_F(\symb{comp})$) returns the sets of all \emph{functional} server
(resp.  client) interfaces of component \symb{comp}.

The predicate \emph{IsExtEnd} checks, for an input
interceptor chain,
whether the first interceptor in the chain is  connected to a
server functional interface of the parent component or, for output
interceptors, whether the last one is connected to a client functional
interface of the parent component.
 Predicate \emph{IsIntEnd} is the symmetric, it checks connection with
 the content. However,
the content is not known for a primitive component; in that case the
interfaces of the content are computed by symmetry of
the external (functional) interfaces of the component.

\begin{table}[!tb]
\begin{center}
  \begin{tabular}{ l | p {12 cm} }
 \hline
\multicolumn{1}{c|}{ Predicate name} & \multicolumn{1}{c}{Predicate Definition} \\ 
 \hline
IsInterc & $\symb{IsInterc}(comp : Comp, ~membr : Membr) \Leftrightarrow$\newline
           \text{}~~$\exists ic.IsIntercChain(ic, membr) \land comp \in ic;$\\
\hline
IsIntercChain & $\symb{IsIntercChain}(\{i_{1}\,...i_{K}\} : \text{set of} ~Comp, ~membr : Membr) \Leftrightarrow$\newline
   $\{i_{1}\,...i_{K}\}  \subseteq Comp(membr) \land \exists SI_1...SI_K, CI_1...CI_K.$\newline
    \text{}~$(\forall k \in \{1...K\}).$
    $(\symb{GetSItf}_F(i_k) \!=\! \{SI_k\}  \land
    \symb{GetCItf}_F(i_k) \!=\! \{CI_k\} ) ~\land $
\newline
    \text{}~$(\forall k \in \{1...K-1\}).( \exists b_k \in \symb{Binding}(\symb{membr}).$\newline
    \text{}\qquad$\symb{GetSrc}(b_k, \symb{membr})=CI_{k} \land \symb{GetDst}(b_k, \symb{membr}) = SI_{k+1}) \land$\newline
%
    \text{}~   $\exists b_0,b_K \in \symb{Binding}(\symb{membr}).
\exists I_0,I_K : \Itf.$\newline
    \text{}\qquad   $\big((\symb{IsExtEnd}(I_0,b_0,SI_1,\symb{membr},I_0) \land \symb{IsIntEnd}(CI_k,b_K,I_K,\symb{membr},I_K))$\newline
     \text{}\qquad $\lor (\symb{IsIntEnd}(I_0,b_0,SI_1,\symb{membr},I_0) \land
\symb{IsExtEnd}(CI_k,b_K,I_K,\symb{membr},I_K)) \big)$\\
%
\hline
IsExtEnd &
  $\symb{IsExtEnd}(\symb{CI}:\symb{CItf},~b:Binding,~\symb{SI}:\symb{SItf},~\symb{membr}:Membr,~I:\Itf) \Leftrightarrow$\newline
   \text{}\quad $\symb{GetSrc}(b, \symb{membr})=\symb{CI} \land \symb{GetDst}(b, \symb{membr})=\symb{SI} \land Nature(I) = \textsf{F}\land $\newline
  \text{}\quad$\symb{Sym(I)}\in\symb{GetItf}(\symb{Parent}(\symb{membr}))$\\
 \hline
IsIntEnd &
  $\symb{IsIntEnd}(CI:\symb{CItf},~b:Binding,~SI:\symb{SItf},~\symb{membr}:Membr,~I:\Itf) \Leftrightarrow$\newline
   \text{}\quad $\symb{GetSrc}(b, \symb{membr})=CI \land \symb{GetDst}(b, \symb{membr})=SI \land Nature(I) = \textsf{F}\land$ \newline
    \text{}\quad $\symb{If}~ \symb{IsComposite} (\symb{Parent}(membr))
   ~then  \newline
   \text{}\quad ~\symb{Sym}(I)\in\symb{GetItf}(\symb{Cont}(\symb{Parent}(\symb{membr}))) $  \newline
\text{}~  $\symb{Else}~\symb{Sym}(I)\in\symb{Sym}(\symb{GetItf}(\symb{Parent}(\symb{membr}))) $  
    \\
 \hline
  \end{tabular}
\end{center}
\caption{Interceptor Predicates}
\label{table-Interc-pred}
\end{table}

\subsection{Well-formed components}
\label{sec-wf}
In this section we define the well-formness rules for GCM components.
First, we introduce the core rules and predicates used for the
definition of well-formness. Then, we focus on the
non-functional aspects.

\begin{table}
\begin{center}
  \begin{tabular}{ l | p {12 cm} }
 \hline
\multicolumn{1}{c|}{ Predicate name} & \multicolumn{1}{c}{Predicate Definition} \\ 
 \hline
 UniqueCompNames &
$UniqueCompNames(\symb{Comp}_i^{i\in I} : \text{set of } Comp) \Leftrightarrow$ \newline
$\forall i, i' \in I. i \neq i' \Rightarrow \symb{CName}(\symb{Comp}_{i}) \neq \symb{CName}(\symb{Comp}_{i'})~~
$\\
 \hline
 
 UniqueItfNames &
$\symb{UniqueItfNames}(\Itf_i^{\,i\in I} :\text{set of } \Itf) \Leftrightarrow$ \newline
$\forall i,i'\in I.\, i\neq i'\Rightarrow Name(\Itf_i)\neq Name(\Itf_{i'})~~
$\\
 \hline

 BindingRoles &
$
\symb{BindingRoles}(b:\symb{Binding}) \Leftrightarrow$ \newline
$\symb{Role}(\symb{GetSrc}(b,\symb{Parent}(b))) = \textsf{C} ~\land \symb{Role}(\symb{GetDst}(b,\symb{Parent}(b))) = \textsf{S}$ \\
 \hline

 BindingTypes &
$
\symb{BindingTypes}(b:\symb{Binding}) \Leftrightarrow $ \newline
$\symb{MSignature}(\symb{GetDst}(b,\symb{Parent}(b))) \leq \symb{MSignature}(\symb{GetDst}(b,\symb{Parent}(b))) $\\
 \hline
CardValidity &
$\symb{CardValidity}(\symb{Binding}_l^{l\in L} : \text{set of } \symb{Binding}) \Leftrightarrow$\newline
$\forall l,l' \in L.\,l \neq l'\Rightarrow$\newline
\text{}\qquad$\symb{GetSrc}(\symb{Binding}_l, \symb{Parent}(\symb{Binding}_l))\!\neq\! \symb{GetSrc}(\symb{Binding}_{l'}, \symb{Parent}(\symb{Binding}_{l'}))$\\
\hline
  \end{tabular}
\end{center}
\caption{Core Predicates}
\label{table-pred}
\end{table}

\subsubsection{Core}
 Table \ref{table-pred} formalizes the auxiliary predicates which are used for the definition of the following  constraints:

\begin{itemize}
\item
Components naming constraint ($UniqueCompNames$): all the components at the same level of hierarchy must have
different names. This restriction is due the fact that components are
referenced by their name; typically, \symb{QName} can be of the form
$\textsf{CName}.\textsf{Name}$; and if two components had the same name
the functions \symb{GetSrc} and \symb{GetDst} would not return a
single deterministic result.
 The two contents of two different composite
components as well as two membranes are considered to be different
name-spaces. A membrane and a content are also different name-spaces
even if they belong to the same component.
\item
Interfaces naming constraint ($\symb{UniqueItfNames}$): all the interfaces of a component must have different
names. This constraint will be checked separately, for the external
interfaces of a component, and for the internal interfaces
 of a content. This constraint also ensures the determinacy of the
 functions \symb{GetSrc} and \symb{GetDst}.
\item Role constraint ($BindingRoles$): a binding must go from a client interface to a server interface. 
The predicate uses a function $\symb{Role}(I)$ that returns $\textsf{C}$
(resp. $\textsf{S}$) if $I$
is a client (resp. server) interface.
\item Typing constraint ($BindingTypes$): a binding must bind interfaces of compatible
  types. The compatibility of interfaces means that for each method of
  a client interface there must exist an adequate corresponding method in the
  server interface. In other words, if a client interfaces is
  connected to a server interface and it wants to call some method,
  then this method must actually exist on the server interface.
In general, a corresponding method does not need to have exactly the
same signature as the one required, but can use any sub-typing or
inheritance pre-order available in the modeling language. We denote
$\leq$ such an order between interface signatures.
\item Cardinality constraint, $CardValidity$, ensures that a client interface
is bound to a single server one.
\end{itemize}

We use the previous constraints to define a well-formness predicate,
denoted $WF$. It is defined recursively on component architecture,
namely on primitive components, on composite components, and on contents.

A GCM primitive component is well-formed if all its interfaces have distinct names and its membrane is well-formed.
\begin{eqnarray}
&&\text{Let} ~prim:Prim = \textsf{CName} <\symb{SItf}_i^{i\in I}, \symb{CItf}_j^{j\in J}, M_k^{k\in K}, \Membrane>; \\ \nonumber
&&WF(prim) \Leftrightarrow   
\symb{UniqueItfNames}(\symb{SItf}_i^{i\in I} \cup \symb{CItf}_j^{j\in J}) \land 
WF(\Membrane)
\end{eqnarray}
A GCM composite component is well-formed if  all its external interfaces have distinct names, its content and its membrane are well-formed.
\begin{eqnarray}
&&\text{Let} ~compos:Compos = \textsf{CName} <\symb{SItf}_i^{i\in I}, \symb{CItf}_j^{j\in J},
\Membrane, \Content>; \\ \nonumber
&&WF(compos) 	\Leftrightarrow   
\symb{UniqueItfNames}(\symb{SItf}_i^{i\in I}\cup \symb{CItf}_j^{j\in J}) \land 
WF(\Membrane) 	\land WF(\Content) 	
\end{eqnarray}
The content of a GCM component is well-formed if all its interfaces have distinct names, all its sub-components have distinct names, all its nested bindings ensure valid cardinality,  all its sub-components are well-formed, the role, type and nature constraints are respected for all its sub-bindings. The nature constraint for the bindings relies on The $BindingNature$ predicates. It is discussed  in Section~\ref{sec:NonFunctional} because it is related to the non-functional aspect.
\begin{eqnarray}
&&\text{Let} ~cont : Cont = <\symb{SItf}_i^{i\in I}, \symb{CItf}_j^{j\in J}, \symb{Comp}_k^{k\in K},  \symb{Binding}_l^{l\in L}>; \\ \nonumber
&&WF(cont) \Leftrightarrow  
\left\{\begin{array}{l}
\symb{UniqueItfNames}(\symb{SItf}_i^{i\in I} \cup \symb{CItf}_j^{j\in J}) \land 
UniqueCompNames(\symb{Comp}_k^{k\in K})  \land \\
CardValidity(\symb{Binding}_l^{l\in L}) \land 
\forall k \in K.  WF(\symb{Comp}_k) \land \\
\forall B\in \symb{Binding}_l^{l\in L}. BindingRoles(B) \land 
BindingTypes(B) \land BindingNature(B) 
\end{array}\right.
\end{eqnarray}
The well-formness of a membrane is only significant for the non-functional aspect, it is defined below.
\subsubsection{Non-functional aspects}
\label{sec:NonFunctional}
\begin{table}[t]
\begin{center}
  \begin{tabular}{l | p {11 cm} }
 \hline
\multicolumn{1}{c|}{Predicate name} & \multicolumn{1}{c}{Predicate Definition} \\
 \hline

 UniqueNamesAndRoles &
$ \symb{UniqueNamesAndRoles}(\Itf_i^{\,i\in I} : \symb{set} ~\symb{of}~ \Itf) \Leftrightarrow $ \newline
$\forall i,i'\in I.(i\neq i'\land \symb{Name}(\Itf_i)=\symb{Name}(\Itf_{i'})) \Rightarrow \symb{Role}(\Itf_i) \neq \symb{Role}(\Itf_{i'})$
\\ \hline

BindingNature &
$BindingNature(b:Binding) \Leftrightarrow $ \newline
$\ControlLevel(GetSrc(b, Parent(b))) =
\ControlLevel(GetDst(b, Parent(b))) = 1 \lor $ \newline
$(\ControlLevel(GetSrc(b, Parent(b))) > 1
\land \ControlLevel(GetDst(b, Parent(b))) > 1)$
\\ \hline

  \end{tabular}
\end{center}
\caption{Non-functional Predicates}
\label{table-pred_nf}
\end{table}
In this section we define the static semantics constraints ensuring
safe composition of the non-functional part of a GCM-based
application. Correctness of a membrane relies on the two predicates defined in Table~\ref{table-pred_nf}:

\begin{itemize}
\item
Interfaces naming constraint ($UniqueNamesAndRoles$): if there are two
interfaces with the same names in a membrane, then they must have
different roles. This is as slight relaxation from the
$\symb{UniqueItfNames}$ rule of the general case: we want to allow
corresponding external/internal interfaces pairs of opposite role to
have the same name. This also ensure
compatibility with the  original 
Fractal model, where internal interfaces were implicitly defined as
the  symmetric
of external ones.

\item
Binding nature constraint($BindingNature$) is a rule imposing the
separation of concerns between functional and non-functional
aspects. We want the functional interfaces to be bound together (only
functional requests will be going through these), and non-functional
interfaces to be connected together as a separate aspect. This is simple to impose in the content of composite components, but a little more tricky in the membrane because of the specific status of interceptors.

The solution is to qualify as functional all the components in a
content and all the interceptors, while all other components in the
membrane are non-functional. Then the interfaces are declared functional
or non-functional. From this we compute for each interface a
control level ranging from 1 to 3, where 1 means functional; 2 and 3 mean non-functional. Then
the compatible interfaces are either both ``1'', or both greater than
``1''.
The $ControlLevel$ function is formally defined as:
\begin{eqnarray}
&&\ControlLevel(X:\symb{Comp}) ::=
\left\{\begin{array}{l} \nonumber
  2, ~if ~Parent(X, context):\Membrane \land ~\lnot IsInterc(X, Parent(X))) \\
  1, ~else
\end{array}\right.
\nonumber\\
&&\ControlLevel(X:\symb{Cont}~|~\symb{Membr}) ::= 1
\nonumber\\
&&\ControlLevel(X:\Itf) ::=
\left\{\begin{array}{l}
  \ControlLevel(Parent(X)), ~if ~\Nature(X) = \textsf{F}
 \\ \ControlLevel(Parent(X)) + 1, ~if ~\Nature(X) = \textsf{NF}
\end{array}\right.
\end{eqnarray}

This constraint on the nature of bindings was already mentioned in
\cite{naoumenko10}, we propose here a formal definition that is simpler and more intuitive.

\end{itemize}
As the last step, we define the well-formness predicate for
membranes. A membrane is well-formed if all its sub-components have
distinct names, the naming constraint is respected for its interfaces,
all its sub-components are well-formed, all its bindings ensure a
valid cardinality, and the role, type, nature constraints are respected for all its sub-bindings,  
\begin{eqnarray}
&&\text{Let} ~membr:Membr = <\symb{Comp}_k^{k\in K},  \symb{Binding}_l^{l\in L}>; \\ \nonumber
&&WF(membr)  \Leftrightarrow\left\{\begin{array}{l}
UniqueCompNames(\symb{Comp}_k^{k\in K})  \land 
UniqueNamesAndRoles(\symb{Itf}(membr)) \land \\
\forall k \in K.  WF(\symb{Comp}_k) \land CardValidity(\symb{Binding}_l^{l\in L}) \land \\
\forall B\in \symb{Binding}_l^{l\in L}. BindingRoles(B) \land 
BindingTypes(B) \land BindingNature(B) 
\end{array}\right.
\end{eqnarray}

This section presented an extension to the definition of well-formed
components defined in previous works
\cite{HKK:FMCO09,gaspar:HLPP13}. More precisely, most of the definitions given in
this section allow us to formalize the notion of well-formed
non-functional features and well-formed interceptors, that have never
been formalized before. The new definition of well-formed components
is the basis for the correct composition of distributed components
with  clear separation of
concerns. 

\subsection{Properties}\label{sec:properties}
The well-formed definition of the preceding section guarantees that,
from an architectural point of view, the specified component assembly
behaves well, both at deployment time and during execution. 
More precisely, the constraints specified above ensure the following
properties:
\begin{description}
\item[Component encapsulation] \emph{Bindings do not cross
  boundaries.} Indeed, \symb{GetSrc} and \symb{GetDst} predicates are
  only defined in the context of the parent component, for example,
  the call to \textsl{GetDst} and \symb{GetSrc} inside  the definition
  of the
  \symb{BindingRoles} predicate ensure that both bound interfaces are
  either internal interfaces of the parent component or external interfaces of
  its sub-components, which guarantees that no binding crosses
  component boundaries.
\item[Deterministic communications] The \symb{CardValidity} predicate
  guarantees that each client interface is bound to a single server
  interface, which guarantees that \emph{each communication between
  components is targeted at a single, well-defined,
  destination}. Section~\ref{sec:NxMCommunications} will introduce
  mutlticast interfaces to express communications with multiple destinations.
\item[Unique naming] Several predicates ensure the uniqueness of
  component or interface names
  in each scope (sub-components of the same component, interfaces of
  the same component, etc.). This restriction is crucial for
  \emph{introspection and modification of the component structure}. For
  example rebinding of interfaces can be easily expressed based on
  component and interface names.
\item[Separation of concerns] The definition of non-functional aspects
  ensure that: 1) \emph{each component has a well-defined nature}: functional
  if it belongs to the content, and non-functional if it belongs to
  the membrane. 2) \emph{each interface has a well-defined control level}, 
   depending on the component it belongs
  to and on the nature of the interface. The nature of components and interfaces is defined by the
  control-level (\symb{CL}) predicate. 3) \emph{Bindings only connect
  together functional} (resp. non-functional) \emph{interfaces}. 4) We clearly
  identify \emph{interceptor components} that are the only structural
  exception to these rules: an interceptor is a  component in the
  membrane that can
  have a single functional client and a single functional server
  interface, but as many non-functional interfaces as necessary.
\end{description}

In order to guarantee that the generated ADL deploys correctly,
i.e. without runtime error, it is sufficient to ensure that (1) each
interface and class the ADL file refers to exists (which is ensured by
the generation process), that (2) no binding exception will be raised
at instantiation time (which is ensured by the well-formedness property and
in particular by the determinacy of communications), and that (3) each
mandatory interface is bound and thus the component system can be
started without error (which is again ensured by the
well-formedness property dealing with bindings); finally (4) the
unique names ensure that the components and interfaces can be
manipulated adequately during instantiation. All those arguments ensure
that \emph{each ADL file generated by our platform deploys correctly},
provided the well-formed property is verified by the system.

The properties verified by well-formed components not only guarantee
that the ADL generated from a well-formed specification deploys
correctly, but also ensure some crucial properties concerning the
runtime semantics (deterministic communications, separation of
concerns, reconfigurability ...).

\section{Extension and Related Work}
\label{sec:rw}
While Section~\ref{sec:formal} focused on the core concepts of a
component model with componentized membranes, we illustrate in this
section how our framework can be used to model other features of GCM
or of other component models. We also compare our approach with
related works.

\subsection{Collective communications}
\label{sec:NxMCommunications}
One of the crucial features of GCM is to enable one-to-many and
many-to-one communications through specific interfaces, namely
gathercast and multicast.

In order to specify such communications let us add a Cardinality
($Card$) field in the specification of the GCM interfaces. The cardinality
can be \emph{singleton}, \emph{multicast} or \emph{gathercast}. For example, interface \emph{M1} at Figure~\ref{fig:gcm} is multicast: it sends requests to several workers at the same time. \emph{G1} is gathercast: it can receive requests from several workers. \emph{S1} and \emph{C1} are singleton interfaces.

These new interfaces modify the definition of cardinality validity.
In particular, the multicast interface allows two bindings to originate
from the same client interface. The \symb{CardValidity} is modified
as follows:
\begin{small}
  \[\begin{array}{@{}l@{}}
    \symb{CardValidity}(\symb{Binding}_l^{l\in L} : \text{set of } Binding) \Leftrightarrow
    \forall \symb{itf}: \Itf.\,\forall l,l' \in L.l \neq l' \\ 
    \quad (\symb{itf}\!=\!GetSrc(Binding_l, Parent(Binding_l))\!=\!GetSrc(Binding_{l'}, Parent(Binding_{l'})
    \Rightarrow Card(\symb{itf}) \!=\! multicast)
  \end{array}
  \]
\end{small}

The intended semantics is that any invocation emitted by a multicast
interface is sent to all the server interfaces bound to it.
Gathercast interface on the contrary synchronizes several invocations
arriving at the same server interface,
they do not entail any structural constraint. An interceptor chain
should not contain any multicast functional interface because it
should transmit invocations in a one-to-one manner.
 
\subsection{Other component models}
Our work is placed in the context of GCM component model but it can
also be used to model the characteristics of most component models, we
review some of them in this paragraph. First, concerning the
functional parts of components, a lot of component models have a
structure similar to GCM components, like, e.g., SCA~\cite{SCAspec} or
Fractal~\cite{fractal:SPE2006}. Our framework can be used to formally
define  architectural correctness for these component
models. Additionally, several frameworks offer different notion of
componentized component control, we focus on these aspects below.

SOFA 2.0~\cite{SOFA:SERA2006} features hierarchical composition and
componentized component control. Interestingly, similarly to our approach,
one of the objectives of SOFA is also to provide formal verification
tools for distributed component systems. 
 It is easy to
adapt our formal specification to SOFA 2.0, more precisely:
\textbf{Micro-controllers} are components dedicated to non-functional
  concerns, they reside in the membrane and are not hierarchical: to
  take them into account, we should \emph{restrict the correctness rules for
  the membrane to only allow  primitive components in the
  membrane}. The other aspects of SOFA are not different from the
model used in this paper. In particular,
 \textbf{delegation chains} are chains of interceptors, following
  exactly the \emph{rules defined  in Section~\ref{sec:interceptors}}. 

 AOKell~\cite{seinturier06aokell} is an extension
of Fractal specifically for componentizing membranes, it is
interesting to note that the authors define a notion of control level,
which is quite similar to the \emph{ControlLevel} function used in our
paper. In this case again, our approach could be used to verify the
correct composition of AOKell-based components, and ensure the safe
design of AOKell component systems. 

In a similar way, Dynaco~\cite{Dynaco05} could
also benefit from our approach, it is a component model targeting
dynamic adaptation and allowing full-fledged components in component's membranes.

\subsection{Related Approaches}

In a slightly different application domain,
BIP~\cite{BensalemBNS10,Basu:2011} is a formal framework that allows
building and analyzing complex component-based systems, both
synchronous (reactive) or asynchronous (distributed) by
coordinating the behaviour of a set of primitive and heterogeneous
components. BIP is supported by a tool set including translators from various
programming languages as Lustre and C into BIP, a compiler for
generating code executable by a dedicated engine, and the
verification tool dFinder.

Concerning the formalization of component structure, the two closest
previous works are the formalization of Fractal in
Alloy~\cite{MERLE:2008}, and the formalization of GCM in
Isabelle/HOL~\cite{HKK:FMCO09}. The first framework focuses on
structural aspects in order to prove the realisability of component
systems, while the second aims at providing lemmas and theorems to
reason on component models at a meta-level, and prove generic
properties of the component model.
 None of those formalizations included
the notion of non-functional components, many-to-many interfaces, or
interceptors. The formal specification of component correctness
defined in this paper could be used to extend the expressiveness of
the component model in  the Alloy and the Isabelle frameworks.
Also the
strength of the present work is that it is directly implemented in the
VCE environment in order to provide tools accessible for the
programmer. The tools presented here ensure, in a graphical
environment, that the component systems composed by the programmer are
well-formed.

From another point of view, in~\cite{gaspar:HLPP13} we investigated a
language for generating correct-by-construction component systems, and
reconfiguring them safely. Similarly to the cases above, this language
does not deal
with the structure of the membrane and one-to-many/many-to-one communications, it should
be extended to the enhanced component structure presented here.

\section{Implementation: the VCE modeling environment}
\label{sec:vce}
\begin{figure}[t]
     \centering
     \includegraphics[width=\textwidth]{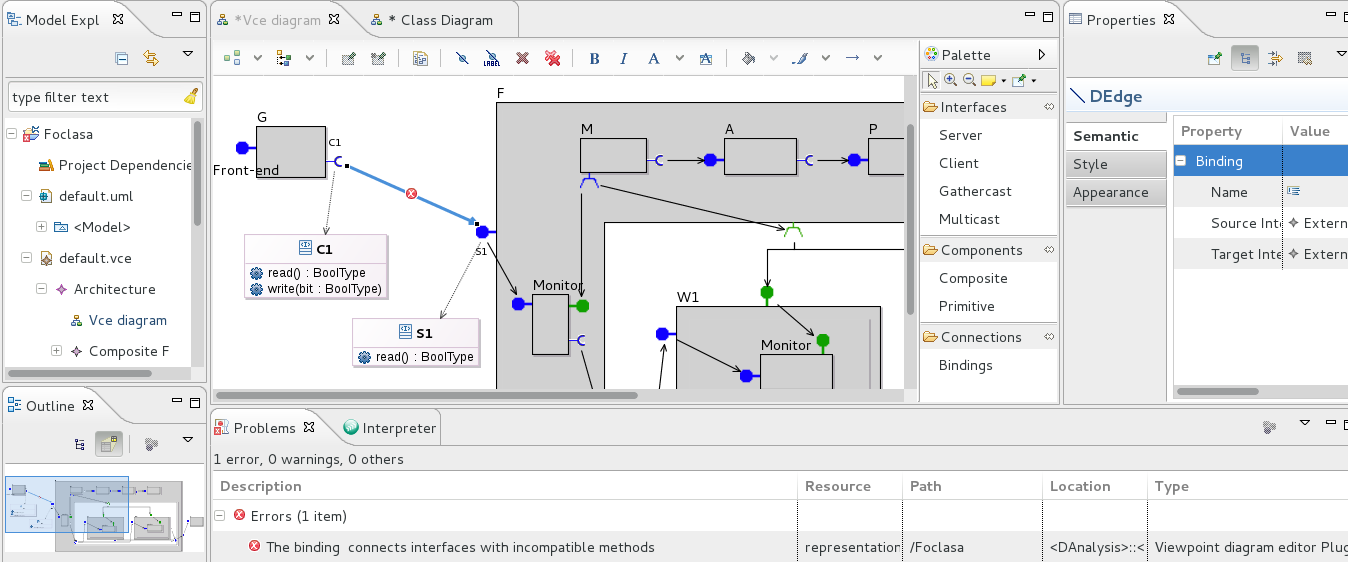}
     \caption{VCE v.3}
     \label{fig:vce_v3}
 \end{figure}

The
 presented formal model and formal constraints were implemented in a tool called VerCors
 Component Editor (VCE v.3) which is the graphical specification part of the VerCors
 project. VCE v.3 combines a set of semantics models and graphical
 diagram editors, the tools for graphical and textual export of models
 and a set of additional wizards. Using VCE v.3 one can design the
 architecture of a componentS-based system, validate its structure
 according to the rules specified in this paper and export its
 representation, either as an ADL (XML) file, that will be used in the
 execution environment, or as a semantic representation, for
 model-checking. We briefly discuss below the basic features
 of the tool.  

VCE v.3 is an Eclipse-based platform built using
ObeoDesigner\footnote{\url{http://www.obeodesigner.com/}},
EMF\footnote{\url{http://www.eclipse.org/modeling/emf/} \qquad \url{http://www.eclipse.org/modeling/gmf/}\label{FN1}} and
GMF\footnotemark[\value{footnote}]. Figure
\ref{fig:vce_v3} is a screen-shot of GCM Components diagram editor. It
has a standard interface of an Eclipse-based application.  

The static semantics constraints, defined in the previous sections are
encoded in VCE v.3 in order to validate component assemblies. The Acceleo language\footnote{http://www.eclipse.org/acceleo/} is used for the definition of such rules. An example of a rule defining the naming constraint for the components is given below:\newline
\centerline{\texttt{[not self.siblings(Component)->collect(name)->includes(self.name)/]}}

This rule will be applied to each component. Here, \texttt{self} is
the component on which the constraint is validated; let us denote it
$C$. The function \texttt{siblings(Component)} returns the set of all the
components in the same container as $C$. The function \texttt{collect(name)}
extracts the names of all such components. Finally,
\texttt{includes(self.name)} checks if there is any component with the same name as $C$ and returns \texttt{true} if it finds one. 

VCE v.3 editor reports about the constraint violations. The
elements of the architecture which are not correct are marked with red
signs. The description of an error is given in a standard Eclipse
\texttt{Problem} panel. Figure~\ref{fig:vce_v3} illustrates an example
of the architecture validation result; the architecture contains only one
error: there is a binding between two interfaces with incompatible
types (\textbf{C1} and \textbf{S1}). 

To conclude, based on the formal definitions given in this paper, we
implemented an environment for the specification of the
component-based applications involving group communications and with
strong separation of functional and non-functional
concerns. The validation of designed models ensures the correctness of the
components assemblies.  
From this specification VerCors environment allows the verification of
the behavioural properties of the system, and the generation of
executable code that is safe by construction and can be run using the
GCM/ProActive library.

\section{Conclusion and Future Work}
\label{sec:concl}
This article presented the formalization of rich component models with
strong separation of functional and non-functional concerns, including
the definition of interceptors. It also presented a tool implementing the formalized model. The main contributions of the article are:
\begin{itemize}
\item
the formal definition of GCM components including non-functional aspect and interceptors;
\item
the formal definition of well-formness predicates including special rules
for  non-functional aspects, binding compatibility, and interceptors;
special care has been put on the separation of concerns: the notion of control level was introduced and formally defined for the interfaces and components;
\item
implementation of a tool for GCM-based systems architecture graphical modeling and their static semantics validation.
\item the identification of basic and crucial properties of the component
  assemblies which are guaranteed by our rules, and by the VCE platform. In particular, 
  when a GCM application specification has been checked valid in VCE, the generated 
ADL file is guaranteed correct by construction, and the corresponding components will 
be built and deployed correctly (with no execution error) by the middleware component factory.
\end{itemize}

One of the key challenges addressed by this paper is the clear
representation of interceptors chains and especially the formalization of predicates which recognize them among the other components inside a membrane.

Our approach is general enough to be extensible in order to take into
account many-to-many communications or other component models.

However, the VerCors platform is not completely implemented yet. We
are now working on the representation and validation of GCM-based
systems behavior using UML State Machines for the
definition of primitive components behavior in VCE v.3. This
functionality is already partially specified and implemented. 
The state-machine and architecture models will be used to generate 
a finite semantic representation, suitable as the input to
verification tools. The structural wellformess rules from this paper
will have to be complemented by others, ensuring the coherency between
the structural and behavioural models.


\bibliographystyle{eptcs}
\bibliography{oasis,internship_report,biblio}
\appendix

\end{document}